\documentclass[epjST]{svjour}
\usepackage{graphics}
\begin{document}
\title{Wigner-localized states in spin-orbit-coupled bosonic ultracold atoms with dipolar interaction}

\author{Y. Yousefi\inst{1} \and E. \"O. Karabulut\inst{1,3} 
\and F. Malet\inst{1,2}  \and J. Cremon\inst{1} \and S. M. Reimann\inst{1} }
\institute{Mathematical Physics, Lund University, LTH, P.O. Box 118, SE-22100 Lund, Sweden
\and Department of Theoretical Chemistry and Amsterdam Center for Multiscale Modeling, 
FEW, Vrije Universiteit, De Boelelaan 1083, 1081HV Amsterdam, The Netherlands
\and Department of Physics, Faculty of Science, Selcuk University,
TR-42075, Konya, Turkey}
\abstract{
We investigate the occurence of Wigner-localization phenomena in bosonic dipolar ultracold 
few-body systems with Rashba-like spin-orbit coupling. We show that the latter strongly enhances 
the effects of the dipole-dipole interactions, allowing to reach the Wigner-localized regime for 
strengths of the dipole moment much smaller than those necessary in the spin-orbit-free case.}
\maketitle
\section{Introduction}
\label{intro}

During the last recent years, one of the most active topics in the field of ultracold atom
gases has been the study of spin-orbit (SO) coupling effects, whose realization was discovered 
to be possible by means of properly designed atom-laser coupling schemes \cite{so_cold}. 
The physics of such SO-coupled ultracold gases is described by a Hamiltonian  
rather similar to the one well known from semiconductor-nanostructure systems, where so-called 
Rashba and Dresselhaus SO effects arise naturally due to the presence of internal net electric
fields \cite{so_qdots}. 
Already the first theoretical studies on SO coupling in ultracold atomic gases \cite{so_cold} 
showed the innovative potential of these systems, opening the door for the investigation 
of a collection of new physical phenomena with no analogue in their solid-state counterparts 
\cite{so_cold,Gal13,so_vortexs}. Research in this field became even more intense with
the recent first experimental realization of a SO-coupled Bose-Einstein condensate \cite{Lin11}. 

The interplay between the SO coupling and the atom-atom interactions is known to give rise to 
novel physical properties \cite{Gal13} in both the strongly-correlated \cite{Dal11} and in the 
weakly-interacting regimes \cite{so_vortexs}. Indeed, it has been shown that the SO coupling can 
enhance interaction effects \cite{Zha12,Zha13}, and this issue is at the focus of the present 
article. In particular, we consider a symmetric Rashba-type SO coupling \cite{rashba}, which is 
the most frequently used form in the literature due to its similarity to that known from 
semiconductor nanostructures. However, it has not yet been experimentally realized due to its 
highly symmetric structure. 

Wigner-localization properties of quasi-one- and quasi-two-dimensional dipolar systems have 
been previously addressed in the absence of SO coupling \cite{wigner1d,wigner2d}. Here, we 
investigate how the presence of SO coupling enhances the Wigner-localization effects 
in few-body dipolar ultracold bosonic systems, finding a clear analogy with what was previously 
observed in semiconductor quantum dots \cite{Rei02} with genuine Rashba coupling and 
electron-electron Coulomb interactions \cite{Cavalli}.

The paper is organized as follows. In Sec. 2 we introduce the model and briefly describe 
the theoretical approach. In Sec. 3 we present our results, and in Sec. 4 we draw conclusions 
and give an outlook.

\section{Model and method}
\label{sec:model}

We consider a few ultracold bosonic atoms with mass $m$ confined in the $xy$-plane by a harmonic 
potential with a strong confinement along the $z$ direction, i.e., 
$V_{\rm trap}(x,y,z) = m (\omega_0^2 (x^2 + y^2) + \omega_z^2 z^2)/2$ with trap frequencies satisfying 
$\omega_z\gg\omega_0$. This effectively freezes the motion of atoms in the $z$ direction, yielding 
a quasi-two-dimensional system. In what follows, we express the lengths in units of $l_{\rm 0}$, 
where $l_0=\sqrt{\hbar/(m\omega_0)}$ is the oscillator length in the $xy-$plane, and energies in 
units of $\hbar\omega_0$.  

The atoms have two internal (pseudo-spin) states that are coupled with a properly designed Raman 
laser scheme, giving rise to SO coupling \cite{Lin11}. Although the form of the coupling, which has been realized 
so far by the current experiments \cite{Gal13,Lin11}, is not purely symmetric, we assume here a symmetric Rashba-type SO 
coupling because of being the most commonly addressed form in the literature \cite{rashba} and its similarity to that 
in semiconductor nanostructures \cite{Cavalli}. The single-particle Hamiltonian of the system then reads
\begin{equation}
H_{\rm sp}=(\frac{1}{2}\mathbf{p}^2 + \frac{1}{2}\mathbf{r}^2) + 
k_{\rm SO}(\hat{\sigma_x}\mathbf{p}_y - \hat{\sigma_y}\mathbf{p}_x) \; ,
\label{eq_Hsp}
\end{equation}
where $\mathbf{p}=(p_x,p_y)$ is the momentum operator, $\mathbf{r}=(x,y)$ is the position vector in 
the $xy-$plane, $k_{\rm SO}$ determines the strength of the SO coupling and the $\hat{\sigma}$'s are 
the 2$\times$2 Pauli spin matrices.

Regarding the two-body interactions, we assume that each atom has a dipole moment ${\bf d}$
and they are oriented, by means of an external field, along the same direction in the $xz-$plane 
forming an angle $\Theta$ with respect to the $x$-axis. In the particular case $\Theta=90^{\circ}$, the dipoles 
are aligned perpendicularly to their plane of motion, and the dipole-dipole interaction (DDI) is isotropic 
and purely repulsive \cite{dipolar,Bru08,Cre10,zinner}. As the dipoles are tilted (i.e., $\Theta$ is decreased) 
the DDI becomes anisotropic with increasing attractive regions. Accordingly, the many-body 
Hamiltonian reads
\begin{equation}
\mathcal{H}= \displaystyle\sum_{i=1}^{N} (H_{\rm sp})_i + 
\frac{1}{2} \displaystyle\sum_{i\neq j}^{N} V_{\rm dip}^{\rm 2D}({\bf r_i}-{\bf r_j}) \; ,
\label{eq_Hmb}
\end{equation}
where the effective dipolar interaction ($V_{\rm dip}^{\rm 2D}$) in the $xy$-plane is given by \cite{Cre10}
\begin{eqnarray}
  \frac{V_{\rm dip}^{\rm 2D}({\bf r})}{\hbar\omega_0}=\frac{D^{2}}{\sqrt{8 \pi}}
  \frac{e^{\tilde r_{\perp}^2}}{l_z^3}\biggl\{(2+4 \tilde r_{\perp}^2)
  K_{0}(\tilde r_{\perp}^2)-4 \tilde r_{\perp}^2 
  K_{1}(\tilde r_{\perp}^2)
\nonumber\\
 +\cos^{2}\Theta \biggl[-(3+4 \tilde r_{\perp}^2) 
 K_{0}(\tilde r_{\perp}^2)+(1+4 \tilde r_{\perp}^2) 
 K_{1}(\tilde r_{\perp}^2)\biggl]
\nonumber\\
 +2\cos^{2}\Theta\cos^{2}\phi\biggl[- 2 \tilde r_{\perp}^2 
 K_{0}(\tilde r_{\perp}^2)+(2 \tilde r_{\perp}^2-1)
 K_{1}(\tilde r_{\perp}^2)
\biggl]\biggl\} \; ,
\label{eq_vdipeff}
\end{eqnarray}
with $l_z=\sqrt{\hbar/m\omega_z}$ being the oscillator length in the $z$-direction, and $\tilde r_{\perp} \equiv r/(2 l_z)$. 
Here, $\phi$ is the angle in cylindrical polar coordinates, $K_0$ and $K_1$ are the zeroth- and first-order modified Bessel 
functions of the second kind. The coupling constant $D$ is defined in terms of the dipole moment of the atoms as  
$D=\frac{d}{\sqrt{4\pi\epsilon_0}}\frac{\sqrt{m}}{\hbar\sqrt{l_0}}$ \cite{Cre10}.

Since in this work we focus on Wigner-localization effects, we only consider the case with $54.7^{\circ}<\Theta\leq 90^{\circ}$ 
for which the three-dimensional form of the interaction is repulsive \cite{Cre10}.

We numerically obtain the eigenstates of the many-body Hamiltonian Eq. (\ref{eq_Hmb}) by means of a 
configuration-interaction (CI) approach \cite{Cre10}. For the problem at hand, instead of the usual choice of 2D 
harmonic oscillator orbitals as basis for expanding the many-body states, it is more convenient to construct a new 
basis by including as well the SO term of the single-particle Hamiltonian. This allows for a more natural description 
of the SO coupling effects in the system and to perform the numerical simulations in a more efficient way. As usually 
done in CI calculations, we truncate the number of basis states to a finite but large enough subset that provides a 
converged solution. Due to the rapid growth of the basis size with the number of particles, we restrict our 
simulations to $N\leq 4$. 

\section{Results}
\label{sec:results}

\begin{figure}
\centering
\resizebox{0.75\columnwidth}{!}{
\includegraphics{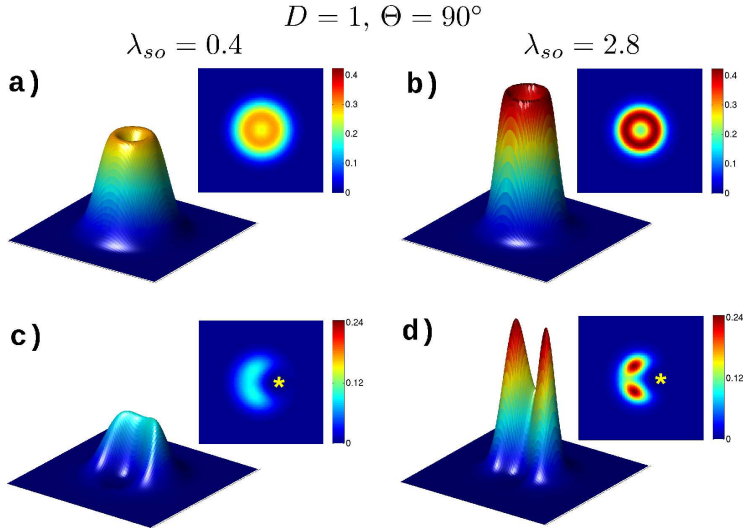} }
\caption{(Color online) Densities (upper row) and pair-correlated densities (lower row) corresponding 
to the ground state of the three-boson system with dipole moment perpendicular to the plane of motion 
($\Theta=90^{\circ}$) and strength $D=1$. The left (a) and c)) and right (b) and d)) columns correspond to 
weak ($\lambda_{SO}=0.4$) and strong ($\lambda_{SO}=2.8$) spin-orbit coupling, respectively. Each contour 
plot, here and in subsequent figures, show the data in the $xy$-plane between the intervals of $-4 < x < 4$ 
(horizontally) and $-4 < y < 4$ (vertically), and the star symbol in the pair-correlated densities refers to 
the position of the reference particle.}
\label{fig:1}       
\end{figure}

\begin{figure}
\centering
\resizebox{0.75\columnwidth}{!}{
  \includegraphics{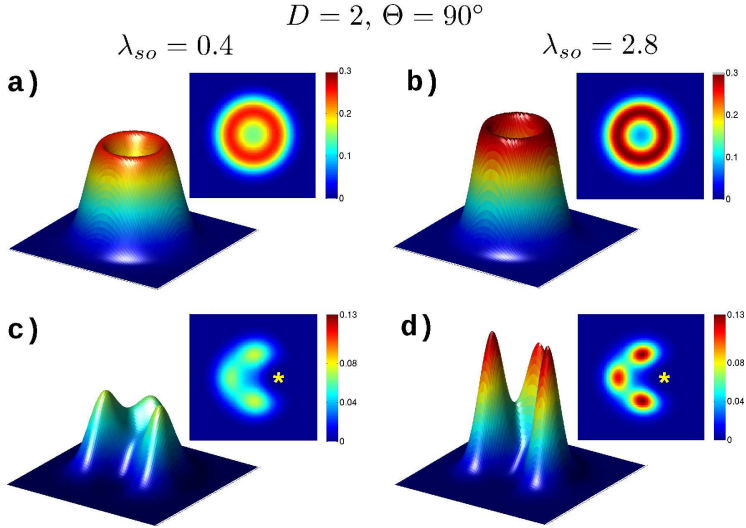} }
\caption{(Color online) Same as Fig.~\ref{fig:1} for $N=4$ and $D=2$.}
\label{fig:2}       
\end{figure}

We first study the case where the dipoles are oriented perpendicularly to the $xy$-plane 
($\Theta=90^{\circ}$). Figs.~\ref{fig:1} and \ref{fig:2} show the densities and pair-correlated densities 
obtained for a system with, respectively, three and four dipoles considering different values of the 
Rashba SO coupling $\lambda_{\rm SO}$ (dimensionless parameter defined by $\lambda_{\rm SO}=l_{\rm 0} k_{\rm SO}$) 
and for a fixed strength of the dipolar interaction $D$. The chosen value for the latter corresponds, in the 
SO-free case, to a moderate interaction strength and to a system in the ``liquid state'', with a rather 
homogeneous density and a pair-correlation density \cite{Cre10} showing no signs of Wigner localization. 
One can see from both figures that this is indeed the case when the SO coupling is small 
(here $\lambda_{\rm SO}=0.4$): the SO term acts only as a small perturbation and the obtained results are 
qualitatively very similar to those corresponding to the case without SO coupling \cite{Cre10}. 

\begin{figure}
\centering
\resizebox{1.0\columnwidth}{!}{
\includegraphics{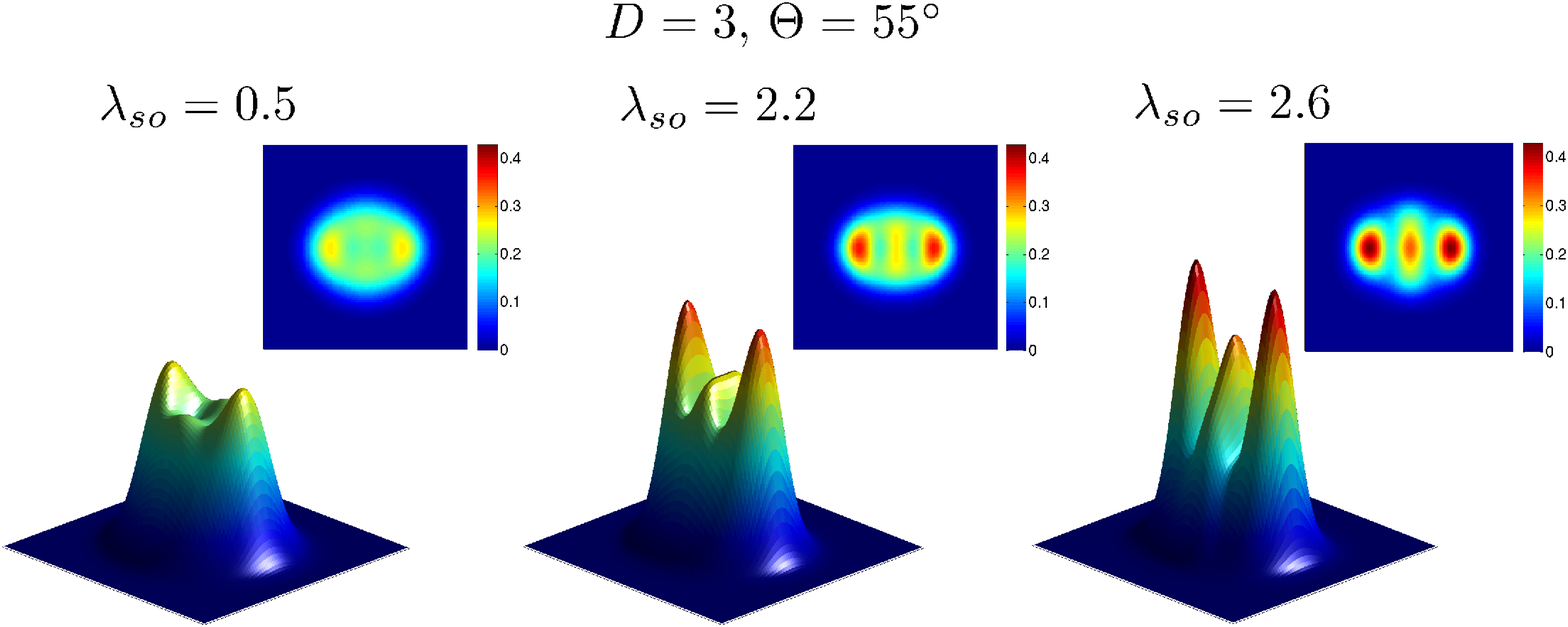} }
\caption{(Color online) Three tilted dipoles with $\Theta=55^{\circ}$, $D=3$ and different SO coupling strengths. 
Each plot has the same color scale, indicated in the figure.}
\label{fig:3}      
\end{figure}

As the strength of the SO coupling is increased, however, the densities (Figs.~\ref{fig:1}b and \ref{fig:2}b) 
start to develop a sharper and more localized structure, forming a ring-shaped profile that reveals a 
strong mutual repulsion. This tendency can also be seen from the corresponding pair-correlation 
densities (Figs.~\ref{fig:1}d and \ref{fig:2}d), which show characteristic Wigner-localized structures 
(triangular and rectangular for $N=$3 and 4, respectively), with a number of clearly marked peaks equal to 
$N-1$ in each case. In the absence of SO coupling, such Wigner-localized configurations are only observed 
for much stronger dipolar interaction strengths (for example, in the case of $N=3$ dipolar bosons similar 
localized states are found for $D=5$ \cite{Cre10}, and here it should be stressed that the interaction strength 
increases as $D^2$). Our results thus clearly show that the presence of SO coupling substantially enhances the effect 
of dipolar interaction in the system, analogously as the effect previously observed in Rashba-coupled semiconductor 
quantum dots with Coulomb interaction \cite{Cavalli}. We have also found that this effect remains in the case of 
tilted dipoles when the interaction becomes anisotropic. Fig.~\ref{fig:3} shows the results corresponding to the 
situation where the dipoles form an angle of $\Theta=55^{\circ}$ with the $x-$axis. In this case the dipolar interaction 
almost vanishes along this direction, favoring the alignment of the particles along it in order to minimize the 
interaction energy. As in the perpendicular case, one clearly observes the strong enhancement of the interaction 
effects as the SO coupling is increased, with a 3-peak Wigner-chain structure emerging when $\lambda_{\rm SO}$ 
becomes large enough. The generation of SO coupling by means of external laser fields appears thus as a very 
effective alternative way to increase the effects of the interactions in ultracold dipolar systems.
\begin{figure}
\centering
\resizebox{0.85\columnwidth}{!}{%
  \includegraphics{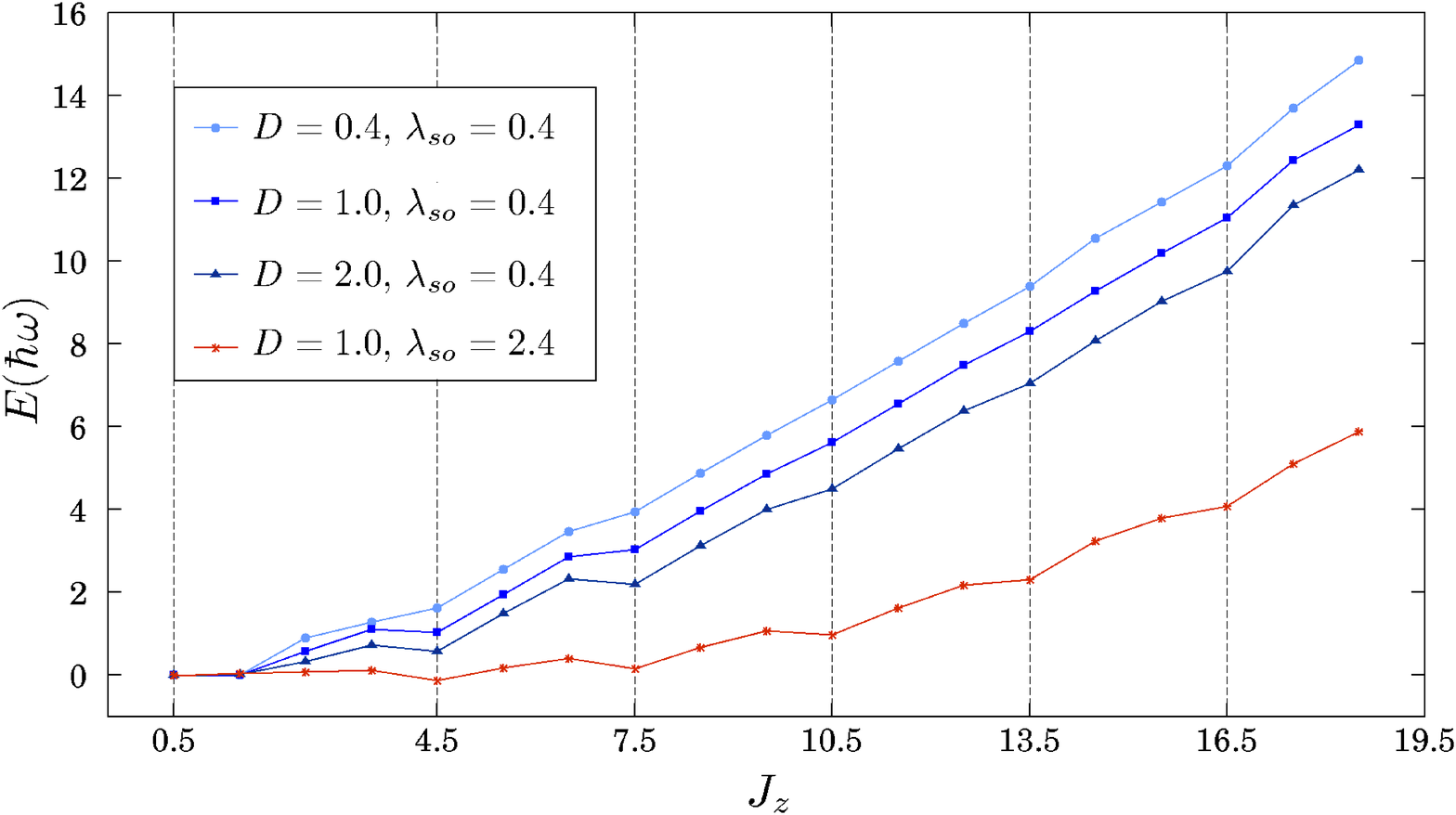} }
\caption{(Color online) Yrast spectrum of three SO-coupled bosons with isotropic and repulsive dipolar interaction.}
\label{fig:4}       
\end{figure}
 
Finally, we want to mention that the analogy between the repulsively interacting SO-coupled dipolar bosons and 
their electronic counterparts is not only limited to the Wigner-localization properties. Fig.~\ref{fig:4} shows the 
yrast spectra (i.e., the ground-state energy as a function of the total angular momentum $J_z$) for three bosons with 
different strengths of the dipolar interaction and SO coupling. When comparing with the results that were obtained 
in the electronic case \cite{Cavalli}, it can be seen that both systems exhibit nearly a flat spectrum in the low 
$J_z$ range due to the almost degenerate single-particle states in this region. However, as the angular momentum 
increases, this flat structure evolves into oscillating yrast lines with a period determined by the number of 
particles in the system. We find that these oscillations become more apparent as the dipoles repel each other 
more strongly (for the isotropic interaction with $\Theta=90^{\circ}$). Moreover, one can see that even when the 
dipolar interaction is weak the periodic behaviour of the spectrum is preserved due to the action of the SO 
coupling, illustrating how the latter enhances the effect of the interactions. Such periodic oscillations in 
the yrast line with recurring local minima are a well-known signature of Wigner localization in electronic systems 
\cite{Cavalli,Maksym}. The above results show how bosonic and fermionic systems become similar in the well-localized 
limit, where the effect of interactions are strong.

\section{Conclusions and outlook}
\label{sec:conclusions}

In this paper we have investigated the interplay between the DDI and the symmetric Rashba SO coupling in a few-body 
bosonic system confined by a harmonic potential. We have shown that the SO coupling strongly enhances the effects of 
the dipolar interaction, giving rise to Wigner-localized structures whose observation would require much larger dipole 
strengths in the spin-orbit-free case. This effect is not only observed in the case of isotropic and purely repulsive 
DDI, where the dipoles are aligned perpendicularly to their plane of motion, but also when they are tilted and the 
interparticle interaction becomes highly anisotropic with both attractive and repulsive regions.

With a setup for their experimental realization having already been proposed \cite{Den12}, dipolar gases with SO 
coupling should be expected to allow the observation of novel interesting phenomena due to the non-trivial interplay 
between the two-body interaction and the Rashba coupling \cite{Wil13}, in particular due to the long-ranged and 
anisotropic nature of the dipole-dipole interaction, as opposed to the more commonly considered contact-type one. 
The high degree of tunability and controllability of these systems should allow for the observation of the 
Wigner-localized states reported in this paper.

Finally, we have restricted our discussion here to the Rashba-type SO coupling, which is the most commonly considered 
one in theoretical studies, but which has not been experimentally realized yet due to its highly symmetric character. 
The effect of an asymmetric coupling, which may model current experiments in a more realistic way will be investigated 
in a future work.    

\textit{Acknowledgements}:
We thank A. Fetter, G. M. Kavoulakis and A. Cavalli for useful discussions. This work was financially supported by the 
Swedish Research Council, the Nanometer Structure Consortium at Lund University, and the POLATOM ESF Research network.

\end{document}